\documentstyle[12pt]{article}

\newcommand{\ton}{\mbox{$\theta_{1}$}}
\newcommand{\ttw}{\mbox{$\theta_{2}$}}
\newcommand{\tonb}{\mbox{$\overline{\theta_{1}}$}}
\newcommand{\ttwb}{\mbox{$\overline{\theta_{2}}$}}

\newcommand{\dg}[1]{\mbox{${#1}^{\dagger}$}}
\newcommand{\hlf}{\mbox{$1\over2$}}

\begin{document}

\baselineskip=14pt plus 0.2pt minus 0.2pt
\lineskip=14pt plus 0.2pt minus 0.2pt

\begin{flushright}
 IUHET 224 \\
 LA-UR-92-2708 \\
August 1992 \\
\end{flushright}

\begin{center}
\Large{\bf
SUPERSQUEEZED STATES} \\

\vspace*{0.25in}

\large
\bigskip

V. Alan Kosteleck\'{y},$^a$ Michael Martin Nieto,$^b$\\
and D. Rodney Truax$^c$

\vspace*{0.2in}
\it

$^a$Department of Physics\\
Indiana University\\
Bloomington, Indiana 47405, U.S.A.\\

\vspace*{0.1in}

$^b$Theoretical Division, Los Alamos National Laboratory\\
University of California\\
Los Alamos, New Mexico 87545, U.S.A. \\

\vspace*{0.1in}

$^c$Department of Chemistry\\
University of Calgary\\
Calgary, Alberta T2N 1N4, Canada\\

\rm
\vspace{0.25in}

\normalsize
%{ABSTRACT}

\end{center}
\begin{quotation}
%\small

We derive the supersqueeze operator
for the supersymmetric harmonic oscillator,
using Baker-Campbell-Hausdorff relations for the
supergroup OSP(2/2).
Combining this
with the previously obtained superdisplacement operator, we derive
the supersqueezed states.  These are the supersymmetric
generalization of the squeezed states of the harmonic oscillator.

\vspace{0.25in}

\noindent PACS: 11.30.Pb, 03.65.-w, 42.50.-p

\end{quotation}

\vspace{2.0in}
Email:  $^a$kostelec@iubacs.bitnet, $^b$mmn@pion.lanl.gov,
$^c$truax@uncamult.bitnet

\newpage

\section{Introduction}

Over the past thirty years
much work has been done on coherent states,$^{1-9}$
especially in the field of quantum optics.
Beyond the harmonic-oscillator system,
coherent states have also been developed for quantum
(Schr\"{o}dinger) systems with general potentials and for general Lie
symmetries.
These states are called (general) minimum-uncertainty coherent
states$^9$ and (general) displacement-operator coherent states.$^{3,6-8}$

There is also a different
generalization of the coherent states of the harmonic oscillator system.
This is the concept of ``squeezed" states.$^{10,11}$
Most notably, squeezed states have been used
in the context of quantum optics and in the context of gravitational wave
detection.
In quantum optics they describe the cases of ``antibunched" and
``bunched" light.$^{12}$
In gravitational wave detection they are used to
describe ``quantum nondemolition" or ``action-back-evading"
measurements.$^{13}$

Recently,
we generalized$^{14}$ the concept of coherent states to
supersymmetric systems,
using the displacement-operator method for supergroups.
This required the use of a general technique for constructing
Baker-Campbell-Hausdorff (BCH) relations$^{15-19}$ for supergroups,$^{20}$
which technique had recently been developed.$^{21-24}$
We discussed three systems:
(i) the super Heisenberg-Weyl algebra,
which defines the supersymmetric harmonic oscillator;
(ii) an electron in a constant magnetic field,
which is a supersymmetric quantum-mechanical system with a
Heisenberg-Weyl algebra plus another bosonic degree of freedom,
and (iii) the electron-monopole system,
which has an OSP(1/2) supersymmetry.
An obvious followup question was whether our supercoherent states
for the harmonic oscillator could be
generalized to supersqueezed states.
It is the purpose of this paper to
demonstrate the positive answer to this question.

In Sec.\ 2 we review the coherent and squeezed states of the harmonic
oscillator from the minimum-uncertainty point of view.  In the following
section we do the same from the displacement-operator point of view,
pointing out the equivalence of the two formulations.$^{10}$
In Sec.\ 4 we review our
harmonic-oscillator supercoherent states.

We start our derivation of the supersqueezed states in Sec.\ 5
by obtaining the differential equations whose solutions
enable us to write the supersqueeze operator as a product of the
exponentials of single algebra elements.
This is done using BCH relations for the supergroup OSP(2/2).
We give the solution of the equations in  Sec.\ 6.
Sec.\ 7 focuses on
the case where only odd operators are involved in the squeeze.
This yields fermionic squeezed states.
Finally, we present the general supersqueezed states.
\newpage

\section{Minimum-uncertainty coherent and squeezed states}

The harmonic-oscillator hamiltonian,
\begin{equation}
H = \frac{1}{2m} p^2 + \frac{1}{2}m\omega^2 x^2,
\end{equation}
is quadratic in the operators $x$ and $p$, which classically vary as
$\sin(\omega t)$ and $\cos(\omega t)$.
The commutation relation of the associated
quantum operators ($\hbar = 1$)
\begin{equation}
[x,p] = i,
\end{equation}
defines an uncertainty relation
\begin{equation}
(\Delta x)^2(\Delta p)^2 \geq 1/4.
\end{equation}

\begin{quotation}
{\it The minimum-uncertainty coherent states for the harmonic oscillator
potential can be defined as those states that minimize the uncertainty
relation (3), subject to the added constraint that the ground state is a member
of the set.}
\end{quotation}

The states that minimize the uncertainty relation (3) are
\begin{equation}
\psi(x) = [2 \pi \sigma^2]^{-1/2}
\exp\left[-\left(\frac{x-x_0}{2 \sigma}\right)^2+ip_0x\right],
\end{equation}
\begin{equation}
\sigma = {\cal S}\sigma_0 = {\cal S}/[2m\omega]^{1/2}.
\end{equation}
When ${\cal S}=1$, these gaussians have the width of the ground
state, so they are the coherent states.
The states are labeled by two parameters, $x_0 = \langle x\rangle $ and $p_0 =
\langle p\rangle $.

The squeezed states of the harmonic oscillator can also be found
from the minimum uncertainty point of view.$^{10}$
In Eqs. (4-5) simply let ${\cal S} \neq 1$.

\begin{quotation}
{\it
The squeezed states of the harmonic oscillator
are minimum-uncertainty gaussians
whose widths are not necessarily that of the ground state.}
\end{quotation}

These states form a continuous
three-parameter set.
Their uncertainty product evolves with time as

\begin{equation}
[\Delta x(t)]^2[\Delta p(t)]^2 = \frac{1}{4}\left[ 1 + \frac{1}{4}
\left({\cal S}^2-\frac{1}{{\cal S}^2}\right)^2\sin^2(2\omega t)\right].
\end{equation}
\newpage

\section{Displacement-operator coherent and squeezed states}

Consider the displacement-operator approach using the oscillator algebra
defined by $a, a^{\dagger}, a^{\dagger}a$, and $I$.
The displacement operator is the unitary
exponentiation of the elements of the factor algebra,
spanned by $a$ and $a^{\dagger}$:
\begin{equation}
D(\alpha) = \exp[\alpha a^{\dagger} - \alpha^* a] =
\exp\left[-\frac{1}{2}|\alpha|^2\right]
\exp[\alpha a^{\dagger}] \exp[-\alpha^* a],
\end{equation}
where the last equality comes from using
a BCH relation.

\begin{quotation}
{\it The displacement operator coherent states of the harmonic oscillator are
obtained
by applying the displacement operator $D(\alpha)$ on  an extremal state,
i.e., the ground state.}
\end{quotation}

Specifically, this yields
\begin{equation}
	D(\alpha)|0\rangle  = \exp[\alpha a^{\dagger} - \alpha^* a] |0\rangle
=\exp\left[-\frac{1}{2}|\alpha|^2\right] \sum_{n} \frac{\alpha ^n}{\sqrt{n!}}
|n\rangle
\equiv |\alpha\rangle  ,
\end{equation}
where $|n\rangle $ are the number states.  With the identifications
\begin{equation}
Re(\alpha)=[m\omega/2]^{1/2}x_0, \hspace{.5in} Im(\alpha)=p_0/[2m\omega]^{1/2},
\end{equation}
these are the same as the minimum-uncertainty coherent states, up to an
irrelevant phase factor.

Obtaining the displacement-operator squeezed states
for the harmonic oscillator
from the coherent states
is more complicated than with the minimum-uncertainty
method. One starts with the ``unitary squeeze operator"
\begin{eqnarray}
S(z) & = & \exp\left[z {\frac{a^{\dagger}a^{\dagger}}{2}}
- z^* {\frac{aa}{2}}\right] \\
& \equiv & \exp\left[G_+ {\frac{a^{\dagger}a^{\dagger}}{2}}\right]
\exp\left[G_0 {\frac{\left(a^{\dagger}a + {\frac{1}{2}}\right)}{2}}\right]
\exp\left[G_- {\frac{aa}{2}}\right] \\
& = &  \exp\left[e^{i\phi}(\tanh r){\frac{a^{\dagger}a^{\dagger}}{2}}\right]
\left({\frac{1}{\cosh r}}\right)^{({\frac{1}{2}}+a^{\dagger}a)}
\exp\left[-e^{-i\phi}(\tanh r){\frac{aa}{2}}\right],
\end{eqnarray}
where Eq. (12) is obtained from a BCH relation$^{25,26}$ and $z \equiv
re^{i\phi}$.  A normal-ordered form for the second term in Eq. (12) is$^{10}$
\begin{equation}
\left({\frac{1}{\cosh r}}\right)^{({\hlf}+a^{\dagger}a)}=
\left({\frac{1}{\cosh r}}\right)^{\hlf}
\left[\sum_{n=0}^{\infty}
{\frac{({\rm sech}\,r - 1)^n}{n!}} (a^{\dagger})^n(a)^n\right].
\end{equation}

The squeezed
states equivalent to the $\psi$ of Eqs. (4-5) are obtained by operating on the
ground state by
\begin{equation}
T(\alpha,z)|0\rangle  = D(\alpha)S(z)|0\rangle  \equiv
|(\alpha,z)\rangle ,
\end{equation}
\begin{equation}
z \equiv re^{i\phi}, \hspace{0.5in} r = \ln {\cal S}.
\end{equation}
Here, $\phi$ is a phase which defines the starting time, $t_0 =
(\phi/2\omega)$, and ${\cal S}$ is the wave-function squeeze of Eq. (5).
Note that $S(z)$ by itself can be considered to be the
displacement operator for the group SU(1,1) defined by
\begin{equation}
K_+ = \frac{1}{2}a^{\dagger}a^{\dagger},\hspace{0.5in} K_- =
\frac{1}{2}aa,\hspace{0.5in}
K_0 = \frac{1}{2}(a^{\dagger}a + \frac{1}{2}),
\end{equation}
so that the $S(z)|0\rangle $ by themselves form SU(1,1) coherent states.
\newpage

\section{Supercoherent states}

The displacement-operator supercoherent states of the harmonic oscillator are
obtained from the super Heisenberg-Weyl algebra, defined by
\begin{equation}
	[a,a^{\dagger}] = I,  \ \ \   \{b,b^{\dagger} \} = I   .
\end{equation}
Using lemma 1 of Ref. 21, one obtains that the superdisplacement operator
is$^{14}$
\begin{eqnarray}
{\bf D}(A,\theta)
& = & \exp[ Aa^{\dagger} -\overline{A}a + \theta b^{\dagger} +
\overline{\theta}b]  \\
& \equiv & D_B(A)\,D_F(\theta),
\end{eqnarray}
where
\begin{equation}
D_B(A) =  \left(\exp[-\frac{1}{2}|A|^2]\exp[A a^{\dagger}]
\exp[-\overline{A} a]\right)
\end{equation}
and
\begin{equation}
D_F(\theta) = \left(\exp[-\frac{1}{2}\overline{\theta}\theta]\exp[\theta
b^{\dagger}]
\exp[\overline{\theta} b]\right).
\end{equation}
The $B$ and $F$ subscripts denote the fact that our supersymmetric
displacement operator can be written as a product
of ``bosonic" and ``fermionic" (more
properly, even and odd) displacement operators.

The variables $\theta$ and $\overline{\theta}$ are Grassmann odd.
They are nilpotent (they only contain a ``soul")
and satisfy anticommutation relations among
themselves and with the fermion operators $b$ and $b^{\dagger}$.
The variables $A$ and $\overline{A}$
are Grassmann even.
Explicit calculation yields

\begin{equation}
	{\bf D}(A,\theta)|0,0\rangle
	= (1 - \hlf\overline{\theta}\theta )|A,0\rangle  + \theta |A,1\rangle  .
\end{equation}

\noindent
The two labels $A$ and $\nu$, with  $\nu = 0,1$, in Eq. (22)
represent the even (bosonic) and odd (fermionic) sectors.
The bosonic state $|A\rangle$
is a superposition of the the number states $|n\rangle$ with the
form of an ordinary coherent state given in Eq. (8).
The fermionic displacement acting alone
produces a Grassmann-valued linear combination
of the states $|0,0\rangle$ and $|0,1\rangle$.
We refer the reader to Ref. 14 for further details
of this construction.
\newpage

\section{Differential equations for the supersqueeze operator}

The previous sections
show that we desire the supersymmetric generalization of
the SU(1,1) squeeze operator of Eqs. (10-12).
The symmetry involved is the supergroup OSP(2/2).
In addition to the su(1,1) algebra elements
of Eq. (16), it has five more:
\begin{eqnarray}
M_0 &=& \frac{1}{2}(b^{\dagger}b-\frac{1}{2}), \nonumber \\
Q_1&=& \frac{1}{2}a^{\dagger}b^{\dagger},  \hspace{.3in}  Q_2 =
\frac{1}{2}ab, \hspace{.3in}
Q_3 = \frac{1}{2}a^{\dagger}b,  \hspace{.3in}  Q_4 = \frac{1}{2}ab^{\dagger}.
\end{eqnarray}

The graded commutation relations among the eight elements
follow from Eq. (17)
(see the Appendix.) Therefore, by using BCH relations for this supergroup,
the supersqueeze operator can in principle
be written as
\begin{equation}
{\bf S}(g)  =  \exp\left[\sum_{i=1}^{6}\alpha_i \hat{g}_i
\right]      = \prod_{i=1}^{8}\exp[\beta_i g_i],
\end{equation}
where $\hat{g}$ is the factor algebra.

We next obtain the differential equations needed to solve Eq. (24),
using the general method developed in Ref. 21.  Consider the
following parametrization of the supersqueeze operator:
\vspace{.25in}
\begin{eqnarray}
{\bf S}(Z,\theta_j,t,\hat{g}) &=& \exp[t(ZK_+ - \overline{Z}K_- +\ton Q_1
     + \tonb Q_2 +  \ttwb Q_3 + \ttw Q_4)] \nonumber \\
\vspace{0.25in}
&=& e^{\gamma_{+}K_{+}}e^{\gamma_{0}K_{0}}e^{\gamma_{-}K_{-}}e^{\beta_{1}Q_{1}}
     e^{\mu M_{0}}e^{\beta_{4}Q_{4}}e^{\beta_{3}Q_{3}}e^{\beta_{2}Q_{2}}
     \nonumber  \\
\vspace{0.25in}
&\equiv & {\bf S_1}(\mu, \gamma_i,\beta_k,t, g).
\end{eqnarray}
Here, the Grassmann-valued variable $Z$
and its complex conjugate $\overline{Z}$
are even,
while the Grassmann-valued
variables $\{ \theta_j \} \equiv \{ \ton , \ttw , \tonb , \ttwb \}$   are odd.
In writing the product form of Eq. (25),
we chose an ordering that yields
relative ease of calculation as well as approximate normal ordering.
Since the fermionic operators are nilpotent,
we ordered them to the right of the purely bosonic operators.
The position of $M_0$ was also chosen for calculational convenience.

Since $\mu$, the $\gamma_j$, and the $\beta_k$ are functions of t,
by taking the derivative of Eq. (25) with respect to t and then
multiplying on the right by ${\bf S}^{-1}$, one has
\begin{equation}
\left[\frac{d}{dt}{\bf S}\right]{\bf S}^{-1}
= \left[\frac{d}{dt}{\bf S_1}\right]{\bf S_1}^{-1} .
\end{equation}
This can explicitly be written as
(a dot over a quantity signifies $\frac{d}{dt}$)
\begin{eqnarray*}
[Z{K}_{+}-\overline{Z}K_{-}&+&\theta_{1}Q_{1}+
\overline{\theta}_{1}Q_{2}+\overline{\theta}_{2}Q_{3}+\theta_{2}Q_{4}] \\
&=& \dot{\gamma}_{+}K_{+} \\
& & +\,[e^{\gamma_{+}K_{+}}]\dot{\gamma}_{0}K_{0}[e^{-\gamma_{+}K_{+}}] \\
& & + \, [e^{\gamma_{+}K_{+}}e^{\gamma_{0}K_{0}}]\dot{\gamma}_{-}K_{-}
     [e^{-\gamma_{0}K_{0}}e^{-\gamma_{+}K_{+}}] \\
& & +\, S_{B}\dot{\beta}_{1}Q_{1}S_{B}^{-1} \\
& & + \,S_{B}[e^{\beta_{1}Q_{1}}]\dot{\mu}M_{0}[e^{-\beta_{1}Q_{!}}]S_{B}^{-1}
\\
& & + \,S_{B}[e^{\beta_{1}Q_{1}}e^{\mu M_{0}}]\dot{\beta}_{4}Q_{4}
     [e^{-\mu M_{0}}e^{-\beta_{1}Q_{1}}]S_{B}^{-1} \\
& & + \,S_{B}[e^{\beta_{1}Q_{1}}e^{\mu M_{0}}e^{\beta_{4}Q_{4}}]
     \dot{\beta}_{3}Q_{3}[e^{-\beta_{4}Q_{4}}e^{-\mu M_{0}}
      e^{-\beta_{1}Q_{1}}]S_{B}^{-1} \\
& & + \,S_{B}[e^{\beta_{1}Q_{1}}e^{\mu
M_{0}}e^{\beta_{4}Q_{4}}e^{\beta_{3}Q_{3}}]
    \dot{\beta}_{2}Q_{2}[e^{-\beta_{3}Q_{3}}e^{-\beta_{4}Q_{4}}
     e^{-\mu M_{0}} e^{-\beta_{1}Q_{1}}]S_{B}^{-1} ,
\end{eqnarray*}
\begin{equation}
\end{equation}
where
\begin{equation}
S_{B}  = [e^{\gamma_{+}K_{+}}e^{\gamma_{0}K_{0}}e^{\gamma_{-}K_{-}}] .
\end{equation}
Note that $S_B$ has a structure analogous to
the ordinary squeeze operator defined in Eq. (10).

All the terms on the right-hand side
of Eq. (27) can be written in nonexponential
form by using BCH formulas and the graded commutation relations.
One has
\begin{eqnarray}
[Z{K}_{+}-\overline{Z}K_{-}&+&\theta_{1}Q_{1}+
\overline{\theta}_{1}Q_{2}
+\overline{\theta}_{2}Q_{3}+\theta_{2}Q_{4}] \nonumber \\
&=& \dot{\gamma}_{+}K_{+} \nonumber \\
& & + [\dot{\gamma}_{0}K_{0} -\dot{\gamma}_{0}{\gamma}_{+}K_{+}]\nonumber \\
& & + \ldots
\end{eqnarray}
The terms left out of the above Eq. (29) become longer and longer, but
can be calculated.
When this is done,
eight equations for the eight variables emerge by extracting the coefficients
of each of the eight generators of osp$(2/2)$; i.e., one equation for each
factor multiplying $K_+, K_0$, etc.
The coefficients yield the following equations:

\begin{eqnarray}
 0 &=& \dot{\mu}+\frac{1}{2}\dot{\beta}_{3}\beta_{4}-
     \frac{1}{2}\dot{\beta}_{2}\beta_{1}e^{-\mu/2} , \\
 Z &= &\dot{\gamma}_{+}-\dot{\gamma}_{0}\gamma_{+}+
     \dot{\gamma}_{-}\gamma^{2}_{+}e^{-\gamma_{0}}
    + \dot{\beta}_{3}\{\frac{1}{2}\beta_{1}e^{-\mu/2}F_{-}^{2}
     -\frac{1}{2}\beta_{4}[\gamma_{+}e^{-\gamma_{0}/2}]F_{-}\}\nonumber \\
 & & + \dot{\beta}_{2}\{-\frac{1}{2}\beta_{1}e^{-\mu/2}
    [\gamma_{+}e^{-\gamma_{0}/2}]F_{-}
    +\frac{1}{2}\beta_{4}[\gamma_{+}e^{-\gamma_0/2}]^{2}\} , \\
 0 &=& \dot{\gamma}_{0}-2\dot{\gamma}_{-}\gamma_{+}e^{-\gamma_{0}}
     +\dot{\beta}_{3}\{\beta_{1}e^{-\mu/2}
     [\gamma_{-}e^{-\gamma_{0}/2}]F_{-}+\frac{\beta_{4}}{2}[1-
     2\gamma_{-}\gamma_{+}e^{-\gamma_{0}}]\} \nonumber\\
&  &+ \dot{\beta}_{2}\{\frac{1}{2}\beta_{1}e^{-\mu/2}
    [1-2{\gamma_{-}\gamma_{+}}e^{-\gamma_{0}}]-\beta_{4}
    [\gamma_{+}e^{-\gamma_{0}}]\} , \\
-\overline{Z} & = & \dot{\gamma}_{-}e^{-\gamma_{0}}
     +\dot{\beta}_{3}\{\frac{1}{2}\beta_{1}e^{-\mu/2}
     [\gamma_{-}^{2}e^{-\gamma_{0}}]
    +\frac{1}{2}\beta_{4}[\gamma_{-}e^{-\gamma_{0}}]\}\nonumber \\
   & & + \dot{\beta}_{2}\{\frac{1}{2}\beta_{1}e^{-\mu/2}
    [\gamma_{-}e^{-\gamma_{0}}]+\frac{1}{2}\beta_{4}[e^{-\gamma_{0}}]\} , \\
\theta_{1} &=& [\dot{\beta}_{1}-\frac{1}{2}\dot{\mu}\beta_{1}+
       \frac{1}{2}\dot{\beta}_{3}\beta_{1}\beta_{4}]F_{-}
    - [\dot{\beta}_{4}e^{\mu/2}
   +\frac{1}{2}\dot{\beta}_{2}\beta_{1}\beta_{4}][\gamma_{+}e^{-\gamma_{0}/2}],
\\
\overline{\theta}_{1} &=& \dot{\beta}_{3}e^{-\mu/2}
     [\gamma_{-}e^{-\gamma_{0}/2}]+\dot{\beta}_{2}e^{-\mu/2}
     [e^{-\gamma_{0}/2}], \\
\overline{\theta}_{2}  &=&
     \dot{\beta}_{3}e^{-\mu /2}F_{-} - \dot{\beta}_{2}e^{-\mu/2}
     [\gamma_{+}e^{-\gamma_{0}/2}], \\
\theta_{2} &=&
     [\dot{\beta}_{1}-\frac{1}{2}\dot{\mu}\beta_{1}+
     \frac{1}{2}\dot{\beta}_{3}\beta_{1}\beta_{4}][\gamma_{-}
     e^{-\gamma_{0}/2}]+[\dot{\beta}_{4}e^{\mu /2}
    +\frac{1}{2}\dot{\beta}_{2}\beta_{1}\beta_{4}]e^{-\gamma_{0}/2} ,
\end{eqnarray}
where
\begin{equation}
F_{\mp}  \equiv  [e^{\gamma_{0}/2}\mp \gamma_{-}\gamma_{+}e^{-\gamma_{0}/2}] .
\end{equation}

Each of the Eqs. (30-37) is linear in time derivatives,
so with some algebra
an equation can be found for each of the eight $t$-derivatives:

\begin{eqnarray}
\dot{\mu}& = &
		   \frac{1}{2}[\overline{\theta}_{1}F_{-}-\gamma_{-}e^{-\gamma_{0}/2}
    \overline{\theta}_{2}]\beta_{1}
    -\frac{1}{2}e^{\mu/2}
   e^{-\gamma_{0}/2}[\gamma_{+}\overline{\theta}_{1}+
    \overline{\theta}_{2}]\beta_{4}, \\
\dot{\gamma}_{+} & = &
    + Z -\overline{Z}\gamma^{2}_{+}-\frac{1}{2}e^{\gamma_{0}/2}
     [\gamma_{+}\overline{\theta}_{1}
     +\overline{\theta}_{2}]\beta_{1}, \\
\dot{\gamma}_{0} & = &
    -2\overline{Z}\gamma_{+}
    -\frac{1}{2}[F_{+}\overline{\theta}_{1}+\gamma_{-}
     e^{-\gamma_{0}/2}\overline{\theta}_{2}]\beta_{1}
    -\frac{1}{2}e^{\mu/2}e^{-\gamma_{0/2}}
    [\gamma_{+}\overline{\theta}_{1}+\overline{\theta}_{2}]\beta_{4}, \\
\dot{\gamma}_{-}& = &-\overline{Z}e^{\gamma_{0}}
    -\frac{1}{2}e^{\gamma_{0}/2}\overline{\theta}_{1}
     [\gamma_{-}\beta_{1}+e^{\mu/2}\beta_{4}], \\
\dot{\beta}_{1} & = &e^{-\gamma_{0}/2}[\theta_{1}+\gamma_{+}\theta_{2}]
      +\frac{1}{4}e^{\mu/2}e^{-\gamma_{0}/2}
      [-\gamma_{+}\overline{\theta}_{1}
    -\overline{\theta}_{2}]\beta_{1}\beta_{4}, \\
\dot{\beta}_{2} & = & e^{\mu /2}[F_{-}\overline{\theta}_{1} -
      \gamma_{-}e^{-\gamma_{0}/2}\overline{\theta}_{2}], \\
\dot{\beta}_{3} & = & e^{\mu/2}e^{-\gamma_{0}/2}
      [\overline{\theta}_{1}\gamma_{+} + \overline{\theta}_{2}], \\
\dot{\beta}_{4} & = & e^{-\mu /2}
    [- \gamma_{-}e^{-\gamma_{0}/2}\theta_{1}+F_{-}\theta_{2}] + \frac{1}{2}

[-F_{-}\overline{\theta}_{1}+\gamma_{-}e^{-\gamma_{0}/2}\overline{\theta}_{2}]
     \beta_{1}\beta_{4} .
\end{eqnarray}
These are the differential equations whose solutions yield the group
parameters for the supersqueezed states.
Note that the boundary conditions needed for these equations
are that the solutions must all be zero when $t=0$.
Then, the supersqueeze operator will be obtained
when we set $t=1$.

\newpage

\section{Solution for the supersqueeze operator}

Equations (39-46)  can be separated into twenty coupled differential equations.
This can be seen by expanding the group parameters in powers of the
odd variables $\theta_j$,
substituting into the eight equations,
and collecting coefficients.
First, the four
even group parameters
$\{ \mu,\,\gamma_+,\,\gamma_0,\, \gamma_- \}$ can each be written as having
three
terms,
containing products of zero, two, or four of the $\theta_j$, respectively.
Second,
the four odd group parameters $\{ \beta_k \}$ can be written as having two
terms,
containing products of one or three  of the $\theta_j$, respectively.
We use a presubscript to denote this; e.g.,
\begin{equation}
\mu = (_0\mu) + (_2\mu)  + (_4\mu), \hspace{0.5in}
\beta_1 = (_1\beta_1) + (_3\beta_1).
\end{equation}
One takes the eight equations (39-46) and expands all of the expressions in
powers of
the $\theta_j$.  The order-zero, -two, and -four pieces of the
even equations are separated and, similarly,
the order-one and -three  pieces of the odd equations are separated.
Note that the lower-order solutions are placed into the higher-order
equations.

To solve the equations, one  first observes that $(_0\mu)=0$.
The equation for $(_0\gamma_+)$
is a Ricatti equation that is solved by the usual procedure,
e.g., as is done to obtain the left-hand term of
the normal squeeze operator S in Eq. (12).$^{25}$
This solution is substituted into the equation for $(_0\gamma_0)$,
whose solution is in turn put into the equation for   $(_0\gamma_-)$.
Except as noted, e.g.,
for $(_0\gamma_+)$ above, the differential equations are all
simple in the sense that there is a  $t$-derivative  of a group
parameter on the left and only powers and hyperbolic functions of $t$ on the
right.
Proceeding, and substituting all previous solutions into subsequent equations,
one directly solves for $(_1\beta_1),\, (_1\beta_2),\, (_1\beta_3),\,
(_1\beta_4),$  and $(_2\mu)$.

The equation for $(_2\gamma_+)$ is an inhomogeneous first-order
equation of the form
\begin{equation}
\dot{q}(t) = k(t)q(t) + f(t).
\end{equation}
Its solution is obtained in a standard way:
\begin{equation}
 q(t) = q^H(t) \int_{0}^{t}\frac{f(\nu)}{q^H(\nu)}d\nu,
\end{equation}
where  $q^H$ is the solution to the homogeneous equation ($f = 0$).

One can then proceed to solve directly for $(_2\gamma_0)$,
$(_2\gamma_-)$,
$(_3\beta_1),\, (_3\beta_2),\, (_3\beta_3),$ $(_3\beta_4),$  and
$(_4\mu)$.
The equations become more complicated, but obtaining the
solutions remains mainly a
question of careful Grassmann-valued algebra.  With
$(_4\gamma_+)$ one has another first-order
inhomogenous differential equation,
whose solution is obtained as above.
Finally, the solutions are completed with
$(_4\gamma_0)$ and  $(_4\gamma_-)$.

In presenting the solutions, we introduce the suggestive notation
\begin{equation}
r \equiv [Z\overline{Z}]^{1/2},  \hspace{1.0in}
e^{i\phi} \equiv [Z/\overline{Z}]^{1/2},
\end{equation}
where $r$ and $e^{i\phi}$ are now understood to represent
Grassmann-valued quantities.
Then, one can make the replacements
\begin{equation}
Z \rightarrow re^{i\phi},  \hspace{1.0in}
\overline{Z} \rightarrow  re^{-i\phi}  .
\end{equation}
Some care is needed because the quantity $e^{i\phi}$ is strictly
defined only for $|\overline{Z}| \ne 0$ and $\overline{z} \ne 0$,
where $\overline{z}$ is the body of $\overline{Z}$.
However, the solutions given below are not affected by this.
We also define
\begin{equation}
c \equiv \cosh y, \hspace{0.5in}
s \equiv \sinh y, \hspace{0.5in} y \equiv rt,
\end{equation}
\begin{equation}
\Phi \equiv \ttwb\ttw\tonb\ton = \ttwb\tonb\ton\ttw  .
\end{equation}
Then, the complete solutions for the group parameters are:
\begin{eqnarray}
\mu & = & 0 \nonumber \\
& & + \frac{1}{2r^{2}}\{[\overline{\theta}_{1}\theta_{1} -
     \overline{\theta}_{2}\theta_{2}](c-1)+
     [\overline{\theta}_{2}\theta_{1}e^{-i\phi}
     -\overline{\theta}_{1}\theta_{2}e^{i\phi}](s-y)\} \nonumber \\
& &+ \frac{\Phi}{r^{4}}\left[c-1-\frac{1}{2}sy\right], \\
\gamma_{+} & = & \left[e^{i\phi}\frac{s}{c}\right] \nonumber \\
&  &  - \frac{e^{i\phi}}{4r^2 c^2} [\overline{\theta}_{1} \theta_{1}(sc-y)
     + e^{i\phi} \overline{\theta}_{1} \theta_{2} (c-1)^{2} \nonumber \\
 & & \hspace{0.5in} + e^{-i\phi} \overline{\theta}_{2} \theta_{1} s^{2}
     + \overline{\theta}_{2} \theta_{2} (sc+ y-2s)] \nonumber \\
& & +  \frac{\Phi e^{i\phi}}{8r^{4}c^{3}} \left[(2y+sy^{2}-s) +
    c(\frac{11}{8}y-2s)+
     \left(-\frac{5}{8}s c^{2} +\frac{1}{4}sc^{4}\right)\right], \\
\gamma_{0} & = &[-2 \ln c]  \nonumber \\
& & + \frac{1}{2r^{2}}[\overline{\theta}_{1}\theta_{1}(\frac{-ys}{c} +c-1)
     + e^{i\phi}\overline{\theta}_{1}\theta_{2}(\frac{s}{c}-s) \nonumber \\
& & \hspace{0.35in} +  e^{-i\phi}\overline{\theta}_{2}\theta_{1}
   (-\frac{s}{c}+s)+
   \overline{\theta}_{2}\theta_{2}(\frac{2+ys}{c}-c-1)] \nonumber \\
& & +{\frac{\Phi}{8r^{4}c^{2}}}[(y^{2}-1-2ys)-c(\frac{11}{4}ys+4)\nonumber \\
& & \hspace{0.5in}   +c^{2}(2{\ln c}+8c-3-4ys - \frac{1}{4}s^{2})], \\
{\gamma}_{-} & = &\left[-e^{-i\phi}\frac{s}{c}\right] \nonumber \\
& & + (\frac{e^{-i\phi}}{4r^{2}c^{2}})
      [\overline{\theta}_{1}\theta_{1}(sc-y)
       -e^{i\phi}\overline{\theta}_{1}\theta_{2}s^{2} \nonumber \\
 & & \hspace{0.5in}-e^{-i\phi}\overline{\theta}_{2}\theta_{1}(c-1)^{2}
       +\overline{\theta}_{2}\theta_{2}(sc+y-2s)] \nonumber \\
& & - \frac{\Phi e^{-i\phi}}{8r^{4}c^{3}}\left[(2y+sy^{2}-s) +
      c(\frac{11}{8}y-2s)+
     \left(sc^{2}(\frac{15}{8}+2\ln c)-\frac{9}{4}c^{3}y\right)\right], \\
\beta_{1}&=&\frac{1}{r}[s\theta_{1}+(c-1)e^{i\phi}\theta_{2}] \nonumber \\
& &+\frac{1}{4r^{3}}[\overline{\theta}_{2}\theta_{1}\theta_{2}(y-2cs+yc)
     +e^{i\phi}\overline{\theta}_{1}\theta_{1}\theta_{2} (2c(1-c)+ys)], \\
\beta_{2}&=&\frac{1}{r}[s\overline{\theta}_{1}
      +(c-1)e^{-i\phi}\overline{\theta}_{2}]\nonumber \\
& &+\frac{1}{4r^{3}}[\overline{\theta}_{2}\theta_{1}\theta_{2}
     (yc-s+\frac{1}{2}(sc-y))
    + \overline{\theta}_{2}\overline{\theta}_{1}\theta_{1}e^{-i\phi}(ys-3(c-1)
    -\frac{1}{2}s^{2})], \\
\beta_{3}&=&\frac{1}{r}[(c-1)e^{i\phi}\overline{\theta}_{1}
      +s\overline{\theta}_{2}]\nonumber \\
& &+\frac{1}{4r^{3}}
    [e^{i\phi}\overline{\theta}_{2}\overline{\theta}_{1}\theta_{1}2(ys-2(c-1))
     + \overline{\theta}_{2}\overline{\theta}_{1}\theta_{1}2(yc-s)], \\
\beta_{4} &=&\frac{1}{r}[(c-1)e^{-i\phi}\theta_{1}+s\theta_{2}] \nonumber \\
 & & +\frac{1}{4r^{3}}
     [e^{-i\phi}\overline{\theta}_{2}\theta_{1}\theta_{2}(-4c^{2}+4c+2ys)
     + \overline{\theta}_{1}\theta_{1}\theta_{2}(-4sc+2s+2yc)] .
\end{eqnarray}

Setting $t = 1$ yields the general supersqueeze group parameters.
In zero and first order,
one notices a symmetry among the parameters.
Compare, for example,
$(_0\gamma_+)$ with $(_0\gamma_-)$
and $(_0\beta_1)$ with $(_1\beta_4)$.
The symmetry remains partial all the way up to fourth order; e.g.,
two of the three components of $(_4\gamma_+)$ and $(_4\gamma_-)$ are
identical.
It is even more evident in the $Z \rightarrow 0$ limit
discussed in the next section.
The symmetry would be modified if the position of group element
$\exp[\mu M_0]$ were different,
say one place to the right or left in
Eq. (25). \newpage

\section{Fermionic squeezed states}

As with the superdisplacement operator, the
supersqueeze operator can be separated into a product of bosonic and
fermionic pieces:
\vspace{.25in}
\begin{equation}
{\bf S}(g) = S_B(g)\,S_F(g) .
\end{equation}
Therefore,  the supersqueezed states are, in general, of
the form
\vspace{.25in}
\begin{eqnarray}
{\bf T}(g) |0,0\rangle  &=& {\bf D}{\bf S}|0,0\rangle  \nonumber \\
&=&D_B(A)D_F(\theta)S_B(g)S_F(g)|0,0\rangle
   = [D_B(A)S_B(g)][D_F(\theta)S_F(g)]|0,0\rangle  \nonumber \\
&\equiv&  T_B(g)T_F(g)|0,0\rangle .
\end{eqnarray}
The general operator produces a linear combination of
states $|n, \nu\rangle$ with arbitrary $n = 0,1,2,\ldots$ and $\nu = 0$
or $1$.

There is an interesting distinction between the superdisplacement and the
supersqueeze operators.  The superdisplacement operator can be written
as
\vspace{.25in}
\begin{equation}
{\bf D}(A,\theta)  =  D_B(A) D_F(\theta).
\end{equation}
Thus, the bosonic displacement operator $D_B$
depends only on the even
Grassmann variables $A$ and $\overline{A}$,
and the fermionic displacement
operator $D_F$ depends only on the odd Grassmann variables $\theta$ and
$\overline{\theta}$.
However, the same is not true of the bosonic and fermionic
squeeze operators.
There, both operators depend on both even and odd
Grassmann variables:
\vspace{.25in}
\begin{equation}
{\bf S}(Z,\theta_j)  =  S_B(Z,\theta_j) S_F(Z,\theta_j).
\end{equation}
This is because the nonzero graded commutation relations of the
supersqueeze algebra mix the even and odd elements of the algebra, something
which does not happen in the case of the
superdisplacement operator for the coherent states.

The distinction can be seen more clearly by taking limits.
In the limit $\theta_j\rightarrow 0$, one is
left with a bosonic squeeze that has form analogous to that
of the ordinary squeeze operator in Eq. (12):
%\vspace{.25in}
\begin{equation}
{\bf S}(Z,0)  =  S_B(Z,0) = S(Z).
\end{equation}
However, when $Z$ goes to zero the situation is quite
different.
For present purposes,
this is equivalent to taking the limit
$r = |Z\overline{Z}|^{1/2} \rightarrow 0$ in Eqs. (54-61). One finds
\begin{eqnarray}
\mu &=& 0  +
     \frac{1}{4}[\overline{\theta}_{1}\theta_{1}
     -\overline{\theta_{2}}\theta_{2}]t^{2}
     -\frac{1}{24}\Phi t^{4}, \\
\gamma_{+} &=& 0  -
      \frac{1}{4}\overline{\theta}_{2}\theta_{1}t^{2} + 0, \\
\gamma_{0} &=& 0  -
    \frac{1}{4}[\overline{\theta}_{1}\theta_{1}
     +\overline{\theta}_{2}\theta_{2}]t^{2}-\frac{1}{16}\Phi t^{4}, \\
\gamma_{-} &=& 0  -
       \frac{1}{4}\overline{\theta}_{1}\theta_{2}t^{2} + 0, \\
\beta_{1} &= & \theta_{1}t  -
       \frac{5}{24}\overline{\theta}_{2}\theta_{1}\theta_{2}t^{3}, \\
\beta_{2} &=&
        \overline{\theta}_{1} t  +
      \frac{1}{6}\overline{\theta}_{2}\overline{\theta}_{1}\theta_{2}t^{3}, \\
\beta_{3} &=&  \overline{\theta}_{2} t  +
      \frac{1}{6}\overline{\theta}_{2}\overline{\theta}_{1}\theta_{1}t^{3}\\
\beta_{4} &=& \theta_{2} t  -
      \frac{1}{3}\overline{\theta}_{1}{\theta}_{1}\theta_{2}t^{3} .
\end{eqnarray}

Setting $t = 1$, one sees that
\vspace{.25in}
\begin{equation}
{\bf S}(0,\theta_j)  =  S_B(0,\theta_j) S_F(0,\theta_j).
\end{equation}
Therefore, a fermionic squeeze ($\theta_j \neq 0$)
is not defined only in terms of the
fermionic squeeze operator $S_F$.
Rather, as shown in Eq. (75), in
addition it has a soul part from the bosonic squeeze operator $S_B$.  A
fermionic squeeze  on $|0,0\rangle$  produces a Grassmann-valued
linear combination of the states
$|0,0\rangle, |1,1\rangle $, and $|2,0\rangle $.
Specifically, from Eqs. (67-74)
one finds
\begin{eqnarray}
{\bf S}(0,\theta_i)|0,0\rangle  = & &
\left[1-\frac{1}{2}\left(\frac{\tonb\ton}{4}\right)
-\frac{1}{12}\left(\frac{\Phi}{16}\right)\right]|0,0\rangle
 + \left[ \frac{\ton}{2}
-\frac{1}{3}\left(\frac{\ttwb\ton\ttw}{8}\right)\right]|1,1\rangle \nonumber \\
  &-& \frac{1}{\sqrt{2}}\left(\frac{\ttwb\ton}{4}\right)|2,0\rangle.
\end{eqnarray}
Note that we have associated a factor $\hlf$ with each $\theta_j$.
This is due to the $\hlf$ in the bivariant elements of the osp$(2/2)$ algebra.

This result suggests the possibility of
an extension of the above states in the context of field theory,
just as the ordinary squeezed states
can be extended and then interpreted as ``two-photon" coherent states$^{27}$.
The squeezing operation involves products of two operators.
Therefore, with the bosonic squeeze turned off ($Z = 0$)
one expects to excite only two-particle states in the field theory,
which, for example, might perhaps be either two photons or one photon and
one photino.$^{28}$

Therefore,  the general fermionic squeezed states,
which are defined in the two
limits $Z \rightarrow 0$ and $A \rightarrow 0$, are
\begin{eqnarray}
{\bf D}(0,\theta){\bf S}(0,\theta_i)|0,0\rangle  = & &
\left[1-\frac{1}{2}\left(\frac{\tonb\ton}{4}\right)
-\frac{1}{12}\left(\frac{\Phi}{16}\right)\right]
\left[\left(1-\frac{1}{2}\overline{\theta}\theta\right)|0,0\rangle
+\theta|0,1\rangle\right] \nonumber \\
 &+& \left[ \frac{\ton}{2}
-\frac{1}{3}\left(\frac{\ttwb\ton\ttw}{8}\right)\right]
\left[\left(1+\frac{1}{2}\overline{\theta}\theta\right)|1,1\rangle
+\overline{\theta}|1,0\rangle\right] \nonumber \\
 &+& \left[- \frac{1}{\sqrt{2}}\left(\frac{\ttwb\ton}{4}\right)\right]
\left[\left(1-\frac{1}{2}\overline{\theta}\theta\right)|2,0\rangle
+\theta|2,1\rangle\right].
\end{eqnarray}
\newpage

\section{General supersqueezed states}

The general supersqueezed states are given by
\begin{eqnarray}
{\bf T}(A,\theta;Z,\theta_j)|0,0\rangle &=&
D_B(A)D_F(\theta)S_B(Z,\theta_j)S_F(Z,\theta_j)|0,0\rangle \nonumber \\
&\equiv& |A,\theta;Z,\theta_j\rangle.
\end{eqnarray}

The structure of these states can be seen as follows.
{}From the definition of $S_F$,
its action  can be split into the action of five separate elements.
Counting from the right,
the first three group elements act as unity, the
fourth just multiplies $|0,0\rangle$ by a constant,
and the fifth yields a linear combination of the
states $|0,0\rangle$ and $|1,1\rangle$.
Similarly,
$S_B$ can be written as the product of three elements.
The first element, $\exp[\gamma_- K_-]$,  acts as unity and the second
element, $\exp[\gamma_0 K_0]$, yields a new
linear combination of  $|0,0\rangle$ and $|1,1\rangle$.
Next, one comes to
\begin{equation}
\exp[\gamma_+ K_+] = \exp[((_2\gamma_+)+(_4\gamma_+))K_+] \exp[(_0\gamma_+)
K_+].
\end{equation}
Expand $\exp[((_2\gamma_+)+(_4\gamma_+))K_+]$ and commute it
until it is to the left of $D_B(A)$, using the relation
\begin{equation}
D(\alpha)f(a^{\dagger},a) = f(a^{\dagger}-\alpha^*,a-\alpha)D(\alpha).
\end{equation}
Write $|1,1\rangle$ as $a^{\dagger}|0,1\rangle$ and similarly commute
$a^{\dagger}$ through $\exp[(_0\gamma_+)K_+]$ and further until it also
is to the left of $D_B(A)$.
Next, commute $D_F(\theta)$ to the right,  expand, and apply it to the linear
combination of states  $|0,0\rangle$ and $|0,1\rangle$.
The result is a new linear combination of
the states $|0,0\rangle$ and $|0,1\rangle$.

The remaining operator, which
multiplies both these states,
is $D_B(A)\exp[(_0\gamma_+)K_+]$.
This operator
acting on a bosonic ground state
has the same effect as $D_B(A)S_B(Z)$.
That is, it produces a squeezed state,
denoted by $|(A,Z)\rangle$, of the form of Eq. (14), but with $\alpha
\rightarrow A$
and $z \rightarrow Z$.  This state, combined with a ket in the fermion space,
yields
$|(A,Z),\nu \rangle$.  Therefore, in this supersymmetric system we have
produced
a linear combination of the states $|(A,Z),0\rangle$ and $|(A,Z),1\rangle$.

Combining all the above together one finds
\begin{eqnarray}
{\bf T}(A, \theta, Z, \theta_{j})|0,0 \rangle
& = & |A,\theta;Z,\theta_j\rangle \nonumber  \\
& = &      \hat{\mu}\Gamma_{-}h_{1}(a^{\dagger})
      \left[(1-\frac{1}{2}\overline{\theta}\theta)
       \,|(A,Z),0\rangle + \theta\,|(A,Z),1\rangle\right] \nonumber \\
&  & +  \hat{\mu}\Gamma_{+}\frac{\beta_{1}}{2}h_{2}(a^{+})
      \left[\overline{\theta}\,|(A, Z), 0\rangle +
      (1+\frac{1}{2}\overline{\theta}\theta)\,|(A, Z), 1\rangle\right],
\end{eqnarray}
where
\begin{eqnarray}
\hat{\mu} & = &1 - \frac{1}{4}[(_2{\mu})+(_4{\mu})] +
\frac{1}{32}(_2\mu)^{2},\\
{\Gamma}_{\pm} & = &
      1 + \frac{(2 \pm 1)}{4}[(_2{\gamma}_0)+(_4{\gamma}_0)]
       + \frac{(2 \pm 1)^{2}}{32}(_2{\gamma}_0)^{2},\\
 h_{1}(a^{\dagger}) &=& 1 + \frac{1}{2}[(_2{\gamma}_+)+(_4{\gamma}_+)]
       (a^{\dagger}-\overline{A})^{2}
       +\frac{1}{8}(_2{\gamma}_+)^{2}(a^{\dagger}-\overline{A})^{4},\\
h_{2}(a^{\dagger}) & = & \frac{(a^{\dagger}-\overline{A})}{c}
      \left[ 1 + \frac{1}{2}(_2{\gamma}_+)
       (a^{\dagger}-\overline{A})^{2}  \right].
\end{eqnarray}

The supersqueezed state of Eq. (81) shows one major similarity to the
supercoherent state of Eq. (22) and one major difference.
It is the same in that it can be written
as a linear combination of factors times a squeezed
or coherent state in the bosonic sector with occupation number
zero or one in the fermionic sector.
It is different	in that for the supersqueezed state the factors
multiplying these states are polynomials
in up to four bosonic creation operators
on the $|(A,Z),\nu\rangle$ states.
Further, in the limit of no fermionic displacement,
$\theta \rightarrow 0$,
$|(A,Z),0\rangle$ is multiplied only by the polynomial $h_1$
of order four,
while $|(A,Z),1\rangle$ is multiplied only by the polynomial
$h_2$ of order three.

In the limits $A\rightarrow 0$ and $Z \rightarrow 0$, the
supersqueezed states of Eq. (81) reduce to
the fermionic squeezed states of Eq. (77),
as expected.

\bigskip\bigskip
\noindent
\large
{\bf Acknowledgments}\\

\normalsize
This research was supported in part by the
United States Department of Energy under contracts
DE-AC02-84ER40125 and  DE-FG02-91ER40661 (VAK),  W-7405-ENG-36 (MMN),
and by the Natural Sciences and Engineering Research Council of Canada (DRT).

\newpage

\noindent
\large
{\bf APPENDIX:  Superalgebras for supercoherent and
 supersqueezed states}\\

\normalsize
The super Heisenberg-Weyl algebra contains the odd generators, $b$
and \dg{b},
the even generators $a$, \dg{a}, and $I$.
They satisfy the nonzero  graded commutation relations:
$$[a,\dg{a}] = I~,~~\{b,\dg{b}\} = I~.\eqno{(A1)}$$
{}From these operators, we will be able to define the supersymmetric
generalization of the squeeze algebra.

The usual squeeze algebra contains the operators
$$K_{+}=\hlf\dg{a}\dg{a}~,
{}~~K_{-}=\hlf aa~,~~K_{0}=\hlf (\dg{a}a+\hlf)~.\eqno{(A2)}$$
That these operators comprise an su(1,1)
Lie algebra can be seen by calculating their commutation relations.
They are
$$[K_{0},K_{\pm}] = {\pm}K_{\pm}~,~~~~[K_{+},K_{-}] = -2K_{0}~.\eqno{(A3)}$$
These are even elements of the supersqueeze algebra.
There is another even operator and it is
defined as
$$M_{0}=\hlf (\dg{b}b-\hlf)~.\eqno{(A4)}$$
The operator $M_{0}$ commutes with the operators $K_{\pm}$ and $K_{0}$.

In addition to the even operators, there are four odd operators which
are defined as follows:
$$Q_{1}=\hlf\dg{a}\dg{b}~,~~~~Q_{2}=\hlf ab~,\eqno{(A5)}$$
$$Q_{3}=\hlf\dg{a}b~,~~~~~Q_{4}=\hlf a\dg{b}~.\eqno{(A6)}$$
These odd operators satisfy a set of anticommutation relations, namely:
$$\{Q_{j},Q_{j}\}=0~,~~~~j=1,\cdots,4~,$$
$$\{Q_{1},Q_{2}\}=\hlf K_{0}-\hlf M_{0}~,~~\{Q_{1},Q_{3}\}=\hlf K_{+}~,$$
$$\{Q_{1},Q_{4}\}=\{Q_{2},Q_{3}\}=0~,~~\{Q_{2},Q_{4}\}=\hlf K_{-}~,$$
$$\{Q_{3},Q_{4}\}=\hlf K_{0}+\hlf M_{0}~.\eqno{(A7)}$$

The remaining commutation relations, between the even and odd elements, are
$$[K_{+},Q_{1}]=0~,~~[K_{+},Q_{2}]=-Q_{3}~,$$
$$[K_{+},Q_{3}]=0~,~~[K_{+},Q_{4}]=-Q_{1}~,$$
$$[K_{-},Q_{1}]=Q_{4}~,~~[K_{-},Q_{2}]=0~,$$
$$[K_{-},Q_{3}]=Q_{2}~,~~[K_{-},Q_{4}]=0~,$$
$$[K_{0},Q_{1}]=\hlf Q_{1}~,~~[K_{0},Q_{2}]=-\hlf Q_{2}~,$$
$$[K_{0},Q_{3}]=\hlf Q_{3}~,~~[K_{0},Q_{4}]=-\hlf Q_{4}~,$$
$$[M_{0},Q_{1}]=\hlf Q_{1}~,~~[M_{0},Q_{2}]=-\hlf Q_{2}~,$$
$$[M_{0},Q_{3}]=-\hlf Q_{3}~,~~[M_{0},Q_{4}]=\hlf Q_{4}~.\eqno{(A8)}$$
{}From the above graded commutation relations we see that our supersqueeze
algebra is the superalgebra osp$(2/2)$.

To complete the formulation of
the symmetry algebra for the supersymmetric oscillator,
we take the semidirect sum of the
super Heisenberg-Weyl algebra with the osp$(2/2)$ superalgebra.
The additional graded commutation
relations among  these elements are
%\vskip .25cm
$$[K_{+},\dg{a}]=0~,~~[K_{+},a]=-\dg{a}~,$$
$$[K_{-},\dg{a}]=a~,~~[K_{-},a]=0~,$$
$$[K_{0},\dg{a}]=\hlf\dg{a}~,~~[K_{0},a]=0~,$$
$$[M_{0},\dg{b}]=\hlf\dg{b}~,~~[M_{0},b]=-\hlf b~,$$
$$[Q_{1},\dg{a}]=0~,~~[Q_{1},a]=-\hlf\dg{b}~,$$
$$\{Q_{1},\dg{b}\}=0~,~~\{Q_{1},b\}=\hlf\dg{a}~,$$
$$[Q_{2},\dg{a}]=\hlf b~,~~[Q_{2},a]=0~,$$
$$\{Q_{2},\dg{b}\}=\hlf a~,~~\{Q_{2},b\}=0~,$$
$$[Q_{3},\dg{a}]=0~,~~[Q_{3},a]=-\hlf b~,$$
$$\{Q_{3},\dg{b}\}=\hlf\dg{a}~,~~\{Q_{3},b\}=0~,$$
$$[Q_{4},\dg{a}]=\hlf\dg{b}~,~~[Q_{4},a]=0~,$$
$$\{Q_{4},\dg{b}\}=0~,~~\{Q_{4},b\}=\hlf a~.\eqno{(A9)}$$

\end{document}